\documentclass[aps,twocolumn,prl,twoside,superscriptaddress,amssymb,amsmath]{revtex4}
\usepackage{graphicx,multirow}
\newcommand{\tr}{{\rm Tr}}
\begin{document}

%\title{Magic cube: the simplest proof of quantum contextuality}
\title{Minimal Kochen-Specker theorem in finite dimensions}
\author{Sixia Yu}
%\email{cqtys@nus.edu.sg}
\affiliation{Centre for Quantum Technologies and Physics Department,
National University of Singapore, 2 Science Drive 3, Singapore
117542}
\affiliation{Hefei National Laboratory for Physical Sciences at
Microscale and Department of Modern Physics, University of Science
and Technology of China, Hefei, Anhui 230026, P.R. China}
\author{C.H. Oh}
%\email{phyohch@nus.edu.sg}
\affiliation{Centre for Quantum Technologies and Physics Department,
National University of Singapore, 2 Science Drive 3, Singapore
117542}

\begin{abstract}
 In \cite{yo} we proved a strengthened Kochen-Specker theorem in 3 dimensions: non-contextual hidden variable (NCHV) models cannot reproduce all the quantum correlations of two compatible observables, which is  a minimal requirement imposed on the NCHV models. Here we shall exclude the NCHV models with this minimal requirement in $d\ge 4$ dimensions by state-independent and experimentally testable inequalities satisfied by all NCHV models that are required to reproduce only the quantum correlations of at most two compatible observables. Furthermore our proofs use the smallest number of rays known so far, e.g., 25 (instead of 31) rays in $5$ dimensions and $5d-2\lfloor d/3\rfloor$ rays in $d\ge 6$  dimensions.

\end{abstract}

\maketitle
Non-contextuality is a typical classical property: the value of an observable revealed by a measurement is predetermined regardless of which compatible observable might be measured alongside. A maximal set of mutually compatible observables defines a context. Kochen and Specker \cite{ks}, and also Bell \cite{bell2},  showed that quantum mechanics (QM) is contextual by excluding all the non-contextual hidden variable (NCHV) models for a quantum system with more than 2 distinguishable states, known as Kochen-Specker (KS) theorem, via some kinds of logical contradictions. Nowadays the quantum contextuality becomes experimentally testable due to various KS inequalities \cite{5,cab1,pit}, inequalities obeyed by all the NCHV models, and has been confirmed in recent experiments on different physical systems \cite{zeil,ustc,yuji, liu,siex,cab2,ex,3}.

Originally KS theorem imposes a rather strong restriction on the NCHV models it excludes \cite{ks}: the partial algebraic structures of compatible observables in QM must be preserved. That is to say the NCHV models it excludes must admit the so-called {\it KS value assignment} in which the value assigned to the product or the sum of two compatible observables is equal to the product or the sum of the values assigned to these two compatible observables.
Indeed the KS theorem is normally proved by finding a finite set of rays, called {\it KS proof}, to which the KS value assignment is impossible. The original KS proof  in $d=3$ dimensions includes 117 rays \cite{ks} which was reduced to 33 rays by Peres \cite{peres} and Sch{\"u}tte \cite{schu} and to 31 rays by Conway and Kochen \cite{31}. In 4 dimensions Cabello, Estebaranz, and Garc\'{\i}a-Alcaine (CEG) discovered a 18-ray proof \cite{18}. KS proofs in higher dimensions can be constructed from those in lower dimensions by using either the Zimba-Penrose method \cite{d2}, which is valid in $d>5$ dimensions, or the CEG method \cite{d1}. There are also many state-dependent KS proofs, e.g., Clifton's 8-ray proof \cite{clif} and the 5-ray proof in 3 dimensions \cite{5}, in which some rays have pre-assigned values.

It was Peres \cite{peres4} who first noticed that the algebraic structures of compatible observables in QM need not to be completely preserved by NCHV models. This type of KS proofs include Mermin-Peres's magic-square proof in 4 dimensions \cite{peres4} and Mermin's pentagon \cite{m2} in 8 dimensions, in which the sum rule is abandoned. It is due to recently discovered KS inequalities that the restriction is lifted (almost) completely. Instead the NCHV models should reproduce the quantum correlations of at least three compatible observables (as in the KS inequalities of Peres-type), which demands that {\it three or more pairwise compatible observables be measured simultaneously}. QM has this property but it may not hold in a most general NCHV model since a context is well defined by pairwise compatibility. Thus a slight trail of algebraic structures of compatible observables in QM is still implicitly required to be preserved by the NCHV models excluded by KS inequalities.

Quite recently \cite{yo} we proved the KS theorem in 3 dimensions with a 13-ray set for which the KS value assignments do exist but fail to reproduce a certain prediction of QM. Based on this set a magic-cube inequality is discovered to exclude all those NCHV models that have only to reproduce the correlation of at most two compatible observables. This is a minimum number of compatible observables whose correlations have to be reproduced since for non-sequential measurements (or expectation of single observable) there does exist a NCHV model \cite{ks}. The question remains open whether there also exist such KS proofs in higher dimensions, given the numerical nature of the derivation of the magic-cube inequality. In this Letter we provide state-independent KS inequalities in all finite dimensions $(d\ge4)$ satisfied by NCHV models that have only to reproduce the correlations of at most two compatible observables. Moreover our proofs involve a smaller number of rays in all finite dimensions $(d\ge 5)$ comparing with the best KS proofs known.

\begin{figure}
\includegraphics[scale=0.91]{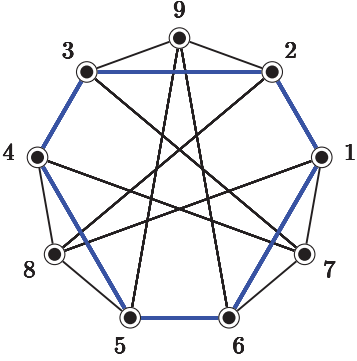}
\caption{(Color online) The orthogonality graph for 18 rays in Eq.(\ref{18r}), which are represented by edges instead of vertices with two rays sharing a common vertex being orthogonal.}
\end{figure}

{\it 18-ray improved KS inequality in 4 dimensions--- }
Motivated by Cabello's 18-ray KS proof \cite{18} we consider the 9-vertex graph shown in  Fig.1. If we can associate each edge $e\in E$, the set of unordered pairs of connected vertices, with a 4-dimensional ray $v_e$ in such a way that
{\it two edges sharing a common vertex are associated with two orthogonal rays}, i.e., two rays $v_e$ and $v_{e^\prime}$ with $e,e^\prime\in E$ being different are orthogonal whenever  $e\cap e^\prime\ne\emptyset$, then we have a KS proof. Indeed the KS value assignment to this 18-ray set amounts to covering all the 9 vertices with disjoint edges. This is impossible because a set of disjoint edges must cover an even number of vertices while we have an odd number of vertices. It turns out that Cabello's 18-ray set in 4 dimensions as presented in \cite{cab1}
\begin{equation}\label{18r}
\begin{array}{c@{\hskip 0.15cm}c@{\hskip 0.15cm}c}
v_{12}=(1,0,0,0)&v_{16}=(0,0,1,\bar1)&v_{34}=(\bar1,1,1,1)\\
v_{18}=(0,1,0,0)& v_{17}=(0,0,1,1)&v_{37}=(1,1,1,\bar1)\\
v_{28}=(0,0,0,1)&v_{67}=(1,\bar1,0,0)&v_{47}=(1,1,\bar1,1)\\
v_{45}=(0,1,0,\bar1)&v_{23}=(0,1,\bar 1,0)&v_{56}=(1,1,1,1)\\
v_{48}=(1,0,1,0)&v_{29}=(0,1,1,0)&v_{59}=(1,\bar1,1,\bar1)\\
v_{58}=(1,0,\bar1,0)&v_{39}=(1,0,0,1)& v_{69}=(1,1,\bar1,\bar1)
\end{array}
\end{equation}
is the unique set that satisfies the aforementioned orthogonality conditions up to a global unitary transformation to all the rays (see Appendix).

The first state-independent experimentally testable KS inequality \cite{cab1} is built upon the above 18-ray set and the quantum correlations of four compatible observables have to be reproduced by the NCHV models it excludes. The KS inequality arising from Peres-Mermin's magic-square proof \cite{cab1} still needs to check the quantum correlations of three compatible observables. In what follows we shall provide a KS inequality in 4 dimensions that involves the correlations of at most two compatible observables. For 18 binary variables $\{\bar v_{e}|e\in E\}$ taking values 0 and 1 the following algebraic inequality holds
\begin{equation}\label{l4}
\bar L_4:=\sum_{e\in E}\bar v_e-\frac12\mathop{\sum_{e\ne e^\prime\in E}}_{e\cap e^\prime\ne\emptyset}\bar v_e\cdot\bar v_{e^\prime}\le 4. \end{equation}
In fact if we denote $\bar v_i=\sum_{e\ni i}\bar v_{e}$ then we can rewrite $\bar L_4=\sum_{i=1}^9\bar v_i(2-\bar v_i)/2$ taking account of the property $\bar v^2_e=\bar v_e$. Because the quadratic function $x(2-x)$ of $x$ reaches its maximal value 1 at $x=1$ and  $\sum_i \bar v_i=2\sum_{e\in E}\bar v_e$ is an even number, not all 9 terms in $\bar L_4$ can reach its maximum. That is to say  for at least one $1\le i\le 9$ it holds $\bar v_i(2-\bar v_i)\le 0$, from which the inequality Eq.(\ref{l4}) follows immediately. Since the last term in Eq.(\ref{l4}) is non-positive we have $\bar L_4\le \bar E:=\sum_{e\in E}\bar v_e$. Also we shall denote by $L_4$ the theory-indendent {\it KS expression} obtained from Eq.(\ref{l4}) with $\bar v$ replaced by $v$.

In any NCHV model all observables have definite values determined only by some hidden variables $\lambda$ that are distributed according to a given probability distribution $\varrho_\lambda$ with normalization $\int d\lambda\varrho_\lambda=1$.  The average of an observable is given by $\langle v\rangle=\int d\lambda \varrho_\lambda v^\lambda$  while the correlation of two observables reads $\langle v\cdot u\rangle=\int d\lambda \varrho_\lambda v^\lambda u^\lambda$. As a result in any NCHV model it holds $\langle L_4\rangle_c:=\int d\lambda\varrho_\lambda  L_4^\lambda\le 4$, where $L_4^\lambda$ is obtained from Eq.(\ref{l4}) by replacing $\bar v$ with corresponding value $v^\lambda$  determined by the hidden variables. Since each correlation is for compatible observables (orthogonal projections) the average $\langle L_4\rangle_q:=\tr(\rho\hat L_4)$ is well-defined in QM for an arbitrary state $\varrho$ of a 4-level system. Here the observable $\hat L_4$ is obtained from Eq.(\ref{l4}) by replacing $\bar v$ with corresponding 1-dimensional projection $\hat v=|v\rangle\langle v|/\langle v|v\rangle$ in which $|v\rangle=v_1|0\rangle+v_2|1\rangle+v_3|2\rangle+\ldots$ for a ray $v=(v_1,v_2,v_3,\ldots)$.
However in in QM it holds the identity $\hat L_4=q_4I_4$ with $q_4=9/2$ and therefore in any state the quantum mechanics predicts $\langle L_4\rangle_q=q_4>4 $.

Furthermore if we require the NCHV models to admit a KS value assignment then we have $\bar v_i=\sum_{e\ni i}\bar v_{e}=1$ for each vertex $i$ and the last term in Eq.(\ref{l4}) vanishes. Choose any triple of unconnected vertices $\{i,j,k\}$, e.g., $\{7,8,9\}$, and denote $ v_{ijk}:=E-v_i-v_j-v_k$, e.g., $v_{789}=v_{12}+ v_{23}+ v_{34}+ v_{45}+ v_{56}+ v_{16}$, which  always form a hexagon (e.g. as shown in Fig.1 as thick blue lines).
Inequality  Eq.(\ref{l4}) becomes now $\bar v_{ijk}\le 1$ so that $\langle v_{ijk}\rangle_c\le 1$, which is referred to as {\it hexagon inequality}. While in QM we have identity $\hat v_{ijk}=3/2$ so that in any state it holds $\langle v_{ijk}\rangle_q=3/2>\langle v_{ijk}\rangle_c$. Thus in order to exclude NCHV models with KS value assignments for 4-level system the hexagon inequality  provides an experimental test that explicitly involves only 6 projections (out of 18). We note that the hexagon inequality and inequality Eq.(\ref{l4}) exclude different kinds of NCHV models.

{\it 25-ray KS inequality in 5 dimensions--- } Consider the orthogonality graph $\Delta_{13}$ shown in Fig.2 as a subgraph on 13 vertices labeled with lowercase letters. As shown in \cite{yo} the graph  $\Delta_{13}$ determines uniquely, up to a global unitary transformation, the 13-ray set $V=y\cup h\cup z$ in 3 dimensions with $z=\{z_k=|k\rangle|k=1,2,3\}$, $y=\{y_k^\sigma=|i\rangle+\sigma |j\rangle|(i,j,k)=1,2,3;\sigma=\pm\}$, and $h=\{h_\alpha=\sum_{k=1}^3(-1)^{\delta_{\alpha k}}|k\rangle|\alpha=0,1,2,3\}$.
Let $\Gamma$ be the adjacency matrix of $\Delta_{13}$, a  $13\times 13$ symmetric matrix with vanishing diagonal elements. And $\Gamma_{uv}=1$ if two vertices $u,v\in V$  are neighbors and $\Gamma_{uv}=0$ otherwise. For 13 binary variables $\{\bar v|v\in V\}$  that take values $0$ and $1$ it holds
\begin{equation}\label{l3}
\bar L_3:=\bar y+\frac{\bar h}2+\bar z-\frac12\sum_{u,v\in V}\Gamma_{uv}\bar u\cdot \bar v\le \frac 72,
\end{equation}
where $\bar h=\sum_{v\in h}\bar v$ for any subset $h\subseteq V$. In fact it is exactly the same inequality as in \cite{yo} if we make a change of variables $a_v=1-2v$. Therefore in any NCHV model it holds $\langle L_3\rangle_c\le c_3:=7/2$ while for all qutrit states it holds $\langle \hat L_3\rangle_q=q_3>c_3$  with $q_3=11/3$ because of the identity $\hat L_3=q_3 I_3$. For later use we note that it always holds $\bar L_3\le \bar V=\bar y+\bar h+\bar z$.

\begin{figure}
\includegraphics[scale=0.9]{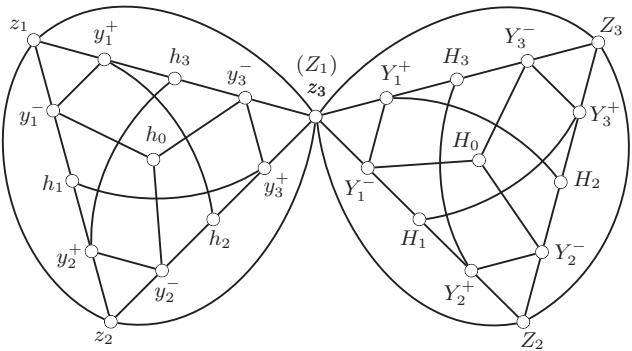}
\caption{ The orthogonality graph for 25 rays in 5 dimensions with rays represented by vertices (hollow dots) and edges, straight or curved, linking two orthogonal rays. For the sake of clarity not all orthogonality relations are shown.}
\end{figure}

According to CEG method \cite{d1} we consider the 25-ray set $V_5=V_+\cup V_-$ in $5$ dimensions with $V_+=\{(v,0,0)|v\in V\}$ and $V_-=\{(0,0,v)|v\in V\}$. For simplicity we shall use the same lowercase letters to represent 5-dimensional rays in $V_+$, e.g., $z_1$ for $(z_1,0,0)$, while the uppercase letters to represent the corresponding rays in $V_-$, e.g., $Z_1$ for $(0,0,z_1)$. Since $V_+\cap V_-=\{z_3=Z_1\}$ we have indeed $|V_5|=25$ rays. Part of the orthogonality relations among these 25 rays are shown in Fig.2. In addition rays labeled with $z_{1,2}$ ($Z_{2,3}$) are orthogonal to all the rays labeled with uppercase (lowercase) letters and rays.

Let $L_3^\pm$ be the corresponding KS expression as defined in Eq.(\ref{l3}) for two 13-ray sets $V_\pm$ and denote $\bar z^\prime=\bar z_{12}+\bar Z_{23}$ where $\bar z_{12}=\bar z_1+\bar z_2$ and $\bar Z_{23}=\bar Z_2+\bar Z_3$. For $2^{25}$ possible values of 25 binary variables $\{\bar v|v\in V_5\}$ taking values 0 or 1 it holds the following inequality
\begin{equation}\label{l5}
\bar L_5:=\bar L_3^++\bar L_3^-+\frac{11}3\bar z^\prime (2-\bar z^\prime)-\bar z_{12}\cdot\bar V_--\bar Z_{23}\cdot\bar V_+\le \frac{43}6,
\end{equation}
where we have denoted $\bar V_\pm=\sum_{v\in V_\pm}\bar v$.
To prove the inequality above we have only to consider the following four cases by noting that $\bar z_{12}$ and $\bar Z_{23}$ are non-negative integers: i) if $\bar z_{12}=\bar Z_{23}=0$ then $\bar z^\prime$ also vanishes so that $\bar L_5\le 2c_3=7$; ii) if $\bar z_{12}=0$ while $\bar Z_{23}\ge1$ then we have $\bar L_5\le q_3+\bar L_3^-\le c_3+q_3=43/6$ since $\bar L_3^+\le \bar V_+$ and $x(2-x)\le 1$; iii) the same bound  holds also in  the case of $\bar z_{12}\ge 1$ while $\bar Z_{23}=0$ for the same reason as case ii); iv) if $\bar z_{12}\ge 1$ and $\bar Z_{23}\ge1$ we have $\bar L_5\le 11/3$ since $\bar L_3^\pm\le \bar V_\pm$ and $x(2-x)\le 1$. If we replace $\bar v$ by corresponding 1-dimensional projection the quantum mechanical prediction reads $\hat L_5=2q_3I_5$ and thus in any state it holds $\langle L_5\rangle_q=22/3>43/6$.

In NCHV models admitting KS value assignments,
it holds  $\bar y+\bar z+3Z_{23}=3$ and $\bar Y+\bar Z+3z_{12}=3$ and  all quadratic terms in Eq.(\ref{l5}) vanish. As a result the KS inequality Eq.(\ref{l5}) now becomes $\bar L_5^\prime:=(\bar h+\bar H)/2+2\bar z^\prime/3\le 7/6$ and thus $\langle L_5^\prime\rangle_c\le 7/6$ must be satisfied by all NCHV models with KS value assignments. While quantum mechanically it holds identity $\hat L_5^\prime=\frac43I_5$, meaning that we have a violation $\langle L^\prime_5\rangle_q=4/3>7/6$ state-independently. Here only 12 projections, namely $\{h_\alpha\}$ and $\{H_\alpha\}$ together with $z_{1,2}$ and $Z_{2,3}$, are involved explicitly.

{\it Improved KS inequality in $d\ge 6$ dimensions---} Consider now $d\ge 6$ and there always exist two nonnegative integers $m,n$ such that $d=3m+4n$. Accordingly the qudit Hilbert space $H_d$ can be decomposed into a direct sum of $m$ qutrit Hilbert spaces $H_3^k$ and $n$ $4-$dimensional Hilbert spaces $H_4^l$ as $H_d=\oplus_{k=1}^mH_3^k\oplus_{l=1}^nH_4^l$.
According to Zimba-Penrose method \cite{d2} we obtain a set $V_d=\cup_k V_k\cup_l E_l$ of $13m+18n$ rays in $d$ dimensions. Here
$V_k$ is the 13-ray set in $d$ dimensions that is supported only on the qutrit Hilbert space $H_3^k$, i.e., obtained from the 13-ray set $V$ by appending necessary zeros, and $E_l$ is the 18-ray set in Eq.(\ref{18r}) that is supported only on the subspace $H_4^l$. By construction, each ray in $V_k$ is orthogonal to all the rays in $V_{k^\prime}$ for $k\ne k^\prime$. Also each ray in $V_k$ is orthogonal to all the rays in $E_l$ and every ray in $E_l$ is orthogonal to all the rays in $E_{l^\prime}$ for $l\ne l^\prime$.

Consider binary variables $\{\bar v|v\in V_d\}$ taking values 0 or 1 and let $L_3^{(k)}$ and $ L_4^{(l)}$ be the KS expressions Eq.(\ref{l3}) for the 13-ray set $V_k$ and  Eq.(\ref{l4}) for the 18-ray set $E_l$, respectively, and $\bar E_l=\sum_{v\in E_l}\bar v$ and $\bar V_k=\sum_{v\in V_k}\bar v$. Then we have the following algebraic inequality
\begin{widetext}
\begin{equation}\label{ld}
\bar L_{d}:=\frac1{q_3}\sum_{k=1}^m{\bar L_3^{(k)}}+\frac1{q_4}\sum_{l=1}^n\bar L_4^{(l)}-\sum_{k=1}^m\bar V_k\cdot\sum_{l=1}^n\bar E_l
-\frac1{q_3}\mathop{\sum_{{k> k^\prime}}^m}_{k,k^\prime=1}\bar V_k\cdot\bar V_{k^\prime}-\frac1{q_4}\mathop{\sum_{l> l^\prime}^n}_{l,l^\prime=1}\bar E_l\cdot\bar E_{l^\prime}\le \frac {c_3}{q_3}.
\end{equation}
\end{widetext}
In fact by noting that $\bar V_T=\sum_k\bar V_k$ and $\bar E_T=\sum_l \bar E_l$ are integers we have only to consider the following four cases: i) $\bar V_T,\bar E_T\ge 1$ ii) $\bar V_T\ge 1$ while $\bar E_T=0$; iii) $\bar E_T\ge 1$ while $\bar V_T=0$; iv) $\bar E_T=\bar V_T=0$.
In the first case where $\bar V_T,\bar E_T\ge 1$ we obtain
$q_3q_4\bar L_d\le 1-(1-q_3\bar E_T)(1-q_4\bar V_T)\le 1$, i.e., $\bar L_d\le 1/(q_3q_4)< c_3/q_3$ by neglecting the last term and taking into account $\bar L_k\le \bar V_k$. In the second case it follows from $\bar E_T=0 $, i.e, $\bar E_l=0$ for all $l=1,2,\ldots,n$ and $\bar V_T\ge 1$ that  $\bar L_4^{(l)}\le \bar E_l=0$ for all $l$. If there is one and only one nonzero $\bar V_k$, say $\bar V_1\ge 1$ and $\bar V_k=0$ for $k\ne 1$, then $\bar L_d\le \bar L_3^{(1)}/q_3\le c_3/q_3$. If there are exactly two nonzero $V_k$'s, say $V_1\ge 1$ and $V_2\ge 1$, we have $q_3\bar L_d\le \bar V_1+\bar V_2-\bar V_1\bar V_2\le 1$ i.e., $\bar L_d\le 1/q_3<c_3/q_3$. If there are more than three nonzero $\bar V_k$'s, say $\bar V_k\ge 1$ for $k=1,2,\ldots, K\ge 3$ then $\sum_{k>k^\prime}\bar V_k\bar V_{k^\prime}\ge \sum_{k}\bar V_{k}\bar V_{k+ 1}\ge \sum_k \bar V_k$ in which we have identified $K+1$ with 1. As a result $\bar L_d\le 0$.  In the third case it follows from $\bar V_T=0$, i.e, $\bar V_k=0$ for all $k=1,2,\ldots,m$ and $\bar E_T\ge 1$, that  $\bar L_3^{(k)}\le \bar V_k=0$ for all $k$. Furthermore if there is one and only one nonzero $\bar E_l$, say $\bar E_1\ge1$, then $\bar L_4^{(l)}\le 0$ except $l=1$ so that $\bar L_d\le L_4^{(1)}/q_4\le 4/q_4< c_3/q_3$. If there are two and only two nonzero $\bar E_l$'s, say $\bar E_1\ge 1$ and $\bar E_2\ge 1$, then  $q_4\bar L_d\le \bar E_1+\bar E_2-\bar E_1\bar E_2\le 1$, i.e., $\bar L_d\le 1/q_4<c_3/q_3$. If there are three or more $\bar E_l\ge 1$ for $l\in K$ then it holds $\sum_{l\in K}\bar E_l\le \sum_{l\ne l^\prime\in K}\bar E_l\bar E_{l^\prime}$ so that $\bar L_d\le 0$. Finally if $\bar E_T=\bar V_T=0$ then $\bar L_d\le 0$. Thus in all cases we have $\bar L_d\le c_3/q_3$. In any NCHV model it therefore holds $\langle L_d\rangle_c\le c_3/q_3$. However in QM we have identity $\hat L_d=I_d$ from which it follows that in any qudit state $\langle L_d\rangle_q=1>c_3/q_3$.

\begin{table}
\begin{equation*}
\renewcommand{\arraystretch}{1.}
 \begin{array}{l@{\hskip 0.4cm}l@{\hskip 0.4cm}l@{\hskip 0.4cm}l}
\hline\hline
d&r_2&r_d\\
3&13&31\\
5&25&29\\
4m  &\displaystyle 18m-2\left\lfloor m/3\right\rfloor&18m          &(m\ge1)   \\
4m+5&\displaystyle 18m+23-2\left\lfloor {(m+2)}/3\right\rfloor&18m+25& (m\ge1)  \\
4m+6&\displaystyle 18m+26-2\left\lfloor {m}/3\right\rfloor&18m+31  & (m\ge 0) \\
4m+7&\displaystyle 18m+31-2\left\lfloor  {(m+1)}/3\right\rfloor&18m+34& (m\ge 0) \\
\hline\hline
\end{array}
\end{equation*}
\caption{A summary of the number $r_2$ of rays used in the improved KS proof in $d$ dimensions in comparison with the smallest known number $r_d$ of rays taken from \cite{d1} in dimensions 5,6, and 7 which are extended to dimensions $4m+\{5,6,7\}$ using Zimba-Penrose method.}
\end{table}

If two integers $m_0,n_0$ is a decomposition of $d=3m_0+4n_0$ then $m=m_0+4l$ and $n=n_0-3l$ is also a decomposition of $d=3m+4n$ for arbitrary $l$ and in this case a number of $13m_0+18n_0-2l$ rays is required. If we choose $l$ optimally, i.e., $m=4\lfloor d/3\rfloor-d$ and $n=d-3\lfloor d/3\rfloor$ where $\lfloor x\rfloor$ denotes the largest integer that is no greater than $x$, then a number $r_2=5d-2\lfloor d/3\rfloor$ of rays is involved. Our results are summarized in Table 1 in which the smallest numbers $r_d$ of rays used in known KS proofs in various dimensions are presented for comparison. For the NCHV models with KS value assignment the inequality
Eq.(\ref{ld}) can be simplified to an inequality that involves $2(d-\lfloor d/3\rfloor)$ projections explicitly.

Our improved KS inequalities derived above, despite of the fact that the properties of discrete nature of binary observables (taking value 0,1) have been used, still hold true for variables taking values between 0 and 1 continuously. In fact all the inequalities considered above involve non-definite quadratic forms of binary variables, which arise from the adjacency matrices of graphs. Therefore the extremal values must be attained on the boundaries which is exactly what have been considered above.

To summarize, we have derived state-independent KS inequalities in $d\ge 4$ dimensions that involve the correlations of at most two compatible observables. And thus NCHV models with only a minimal requirement, in which three or more pairwise compatible observables may even not be simultaneously measurable, cannot reproduce the quantum mechanical predictions on the correlations of two compatible observables, which can be called suitably as a minimal KS theorem. Moreover our improved KS inequalities involve a smaller number of rays than all the KS proofs previously known and therefore are more accessible to experimental tests.

{\it Acknowledgement--- } This work is supported by National
Research Foundation and Ministry of Education, Singapore (Grant No.
WBS: R-710-000-008-271) and NSF of China (Grant No. 11075227).

\setcounter{equation}{0}
\renewcommand{\theequation}{A.\arabic{equation}}

\vskip 0.5cm
{\it Appendix--- }
We  shall prove in what follows that up to a global unitary transformation the orthogonality graph in Fig.1 uniquely determines the 18-ray set. The orthogonality conditions read: {\it two rays labeled with two edges with a common vertex should be orthogonal}. Without loss of generality we can assume
\begin{equation}
\begin{array}{c@{\hskip 0.1cm}cc}
v_{12}=(1,0,0,0)&\multirow{2}{*}{$\Big\}\ v_{28}=(0,0,b,\bar 1)$}\\
v_{18}=(0,1,0,0)\\
v_{17}=(0,0,1,0)&\multirow{2}{*}{$\Big\}\ v_{67}=(1,\bar a,0,0)$}\\
v_{16}=(0,0,0,1)
\end{array}
\end{equation}
with $a,b>0$ because i) any quadruplet of rays sharing a single vertex, e.g., vertex 1, should form a basis; ii)  $v_{28}$ should be orthogonal to both $v_{12}$ and $v_{18}$; iii) complex numbers  $a$ and $ b$ can be chosen to be positive by multiplying a suitable phase fact to $v_{18}$ and $v_{17}$ respectively.
Taking into account of quadruplets of rays having common vertices 2,8,7, and 6 we have
\begin{equation}
\begin{array}{l@{\hskip 0.1cm}l}
v_{29}=(0,u_2,1,b)       & \multirow{2}{*}{$\Big\}\ v_{39}=(g,0,b,\bar 1)$}       \\
v_{23}=(0,u_2^\prime,1,b)&\\
v_{48}=(u_8,0,1,b)       & \multirow{2}{*}{$\Big\}\ v_{45}=(0,h,b,\bar 1)$}\\
v_{58}=(u_8^\prime,0,1,b)                               \\
v_{37}=(a,1,0,u_7)            &\multirow{2}{*}{$\Big\}\ v_{34}=(1,\bar a,s,0)$}            \\
v_{47}=(a,1,0,u_7^\prime)                                       \\
v_{56}=(a,1,u_6,0)            & \multirow{2}{*}{$\Big\}\ v_{59}=(1,\bar a,0,t)$}    \\
v_{69}=(a,1,u_6^\prime,0)
\end{array}
\end{equation}
for some nonzero complex numbers $g,h,s,t$ and $u_k,u_k^\prime$ satisfying  $-u_k^\prime u_k^*=1+a^2$ for $k=6,7$ and $-u_k^\prime u_k^*=1+b^2$ for $k=2,8$.
Furthermore we can choose $t$ to be real and $t>0$ by multiplying the same suitable phase factor to $v_{12}$ and $v_{18}$. From the orthogonality of two quadruplets of rays sharing the vertex 3 and 9
\begin{equation}
\begin{array}{l@{\hskip 0.5cm}ll}
v_{23}=(0,u_2^\prime,1,b)&v_{45}=(0,h,b,\bar 1)\\
v_{34}=(1,\bar a,s,0)         &v_{56}=(a,1,u_6,0)\\
v_{37}=(a,1,0,u_7)            &v_{58}=(u_8^\prime,0,1,b)  \\
v_{39}=(g,0,b,\bar 1)              &v_{59}=(1,\bar a,0,t)
\end{array}
\end{equation}
it follows immediately
\begin{eqnarray}
&\displaystyle a=\frac {s^*}{u_2^\prime}=\frac{u_7}{g^*}=-\frac{t} h=-\frac{u_6^*}{u_8^\prime},&\\
&\displaystyle b=-\frac {u_2^\prime}{u_7^*}=-\frac{g^*}{s}=-\frac{h^*}{u_6}=-\frac{u_8^\prime}{t}.&
\end{eqnarray}
As one result we obtain, keeping in mind that $a,b$ are real, $ab=-s/u_7=-u_7/s$ and thus $ab=1$. From the orthogonality of two quadruplets of rays sharing the vertex 4 and 5
\begin{equation}
\begin{array}{l@{\hskip 0.5cm}ll}
v_{29}=(0,u_2,1,b)  & v_{34}=(1,\bar a,s,0) \\
v_{39}=(g,0,b,\bar 1)         &  v_{45}=(0,h,b,\bar 1)\\
v_{59}=(1,\bar a,0,t)    &  v_{47}=(a,1,0,u_7^\prime)      \\
v_{69}=(a,1,u_6^\prime,0)&  v_{48}=(u_8,0,1,b)
\end{array}
\end{equation}
it follows  $u_6^\prime=-u_2^*$, $g=t$, and $u_8=-s^*$, $u_7^\prime=h^*$ together with
\begin{equation}
\frac ab=\frac t{u_2}=-\frac{u_6^\prime}{g^*}=-\frac{u_7^\prime}{u_8^*}=\frac{s^*}h.
\end{equation}
Consequently we obtain $a/b=t/u_2=u_2/t$ so that $a/b=1$ and thus $a=b=1$. As a result $h=s=-t$ and $u_k$ are all real with $u_k=-u_k^\prime $ and
$u_2=u_7=u_6=u_8=t$ and therefore $t^2=2$. A rotation,
$(1,0)\mapsto (1,1)/\sqrt2$ and $(0,1)\mapsto (1,\bar1)/\sqrt2$ in the subspace spanned by $v_{16}$ and $v_{17}$ recovers Cabello's 18-ray set Eq.(\ref{18r}) exactly.

\newpage
\end{document}